\documentclass[twocolumn,showpacs,showkeys,preprintnumbers,aps,prl,amsmath,amssymb,hyperref,superscriptaddress]{revtex4} 

\usepackage{graphicx}
\usepackage{dcolumn}
\usepackage{bm}

\usepackage{enumerate}
\newcommand\bx{{\mathbf x}}
\newcommand\by{{\mathbf y}}
\newcommand\bz{{\mathbf z}}
\newcommand\bq{{\mathbf q}}
\newcommand\bp{{\mathbf p}}
\newcommand\be{{\mathbf e}}

\newcommand\bk{{\mathbf k}}
\newcommand\bn{{\mathbf n}}

\begin{document}


\title{%
Momentum Conserving Model with Anomalous Thermal Conductivity in Low
Dimensional Systems}

\author{Giada Basile}
\affiliation{Dip. Matematica,
Universit\`a di Firenze,
Viale Morgagni 67a,
50134 Firenze, Italy.}
\author{C\'edric Bernardin}
\affiliation{%
UMPA, UMR-CNRS 5669, ENS-Lyon,
46, all\'ee d'Italie, 69364 Lyon Cedex 07, France.
}%
\author{Stefano Olla}
\email{olla@ceremade.dauphine.fr}
\affiliation{Ceremade, UMR-CNRS 7534, Universit\'e de Paris Dauphine,
  75775 Paris Cedex 16, France.
}%


\begin{abstract}
Anomalous large thermal conductivity has been observed numerically and
experimentally in one- and two-dimensional systems. 
There is an open debate about the role of conservation
of momentum. We introduce a model whose thermal conductivity diverges 
 in dimension 1 and 2 if momentum is conserved, 
while it remains finite in dimension $d\ge 3$. 
We consider a system of harmonic oscillators perturbed by
a non-linear stochastic dynamics conserving  momentum and energy.
We compute explicitly the time correlation function of the energy current
$C_J(t)$, and we find that it behaves, for large time, like $t^{-d/2}$
in the unpinned cases, and like  $t^{-d/2-1}$ when an on site
harmonic potential is present. 
 This result clarifies the role of conservation of
momentum in the anomalous thermal conductivity in low dimensions.
\end{abstract}

\pacs{44.10+i,
      05.60-k,
      63.10.+a,
      66.70.+f          }
\keywords{Thermal conductivity,  Green-Kubo formula, anomalous heat
  transport, Fourier's law, 
non-equilibrium systems}
\maketitle

When a small gradient of temperature $\nabla T$
 is applied to a material, we expect that, in the steady state, the
heat current satisfies Fourier's law
\begin{equation*}
  <J> = -\kappa \nabla T 
\end{equation*}
where $\kappa$ is the conductivity of the material. 
There has been interest in the question of validity of Fourier's
law for low dimensional systems ($d\le 2$), where standard arguments
based on linear response break down  (cf. \cite{blr, sll}
for a review on the subject). 
Anomalous large conductivity is observed experimentally in
carbon nanotubes and numerically in
Fermi-Pasta-Ulam (FPU) systems without pinning (on-site potential),
where numerical evidence shows a conductivity diverging like
$N^\alpha$, with $\alpha<1$ in one dimension, and like $\log N$ 
in dimension 2 (cf. \cite{sll} and references therein). If some non-linearity 
is present in the interaction, finite
conductivity is observed numerically in all pinned cases, and it is
expected in all 3 dimensional cases \cite{blr, sll}.   
Consequently it has been suggested that conservation
of momentum is an important ingredient for the anomalous conductivity
in low dimension \cite{nr}.


In insulating crystals heat is transported by lattice vibrations, and
since the pioneering work of Debye, systems of coupled anharmonic
oscillators have been used as microscopic models for heat
conduction. Non-linearity is extremely important. In fact, in the
linear case the average energy current $<J>$ is independent of the length $N$
of the system, i.e. the conductivity $\kappa_N$ diverges like $N$  \cite{rll}. 
In fact in the harmonic crystal the normal modes of the vibrations
(phonons) do not interact and follow ballistic motion. A finite
asymptotic conductivity instead should result eventually by the diffusive
behavior of phonons due to phonon-phonon interaction caused by
anharmonicity.

Since conductivity in non-linear systems is difficult to compute or
estimate analytically, it is natural to model the nonlinearities by
stochastic perturbations of the linear dynamics. In some sense these
stochastic perturbations simulate (qualitatively) the long time
(chaotic) effect of the deterministic non-linear model.

We study in this Letter a stochastic model 
where conductivity can be explicitly computed,
and diverges in dimension 1 and 2 when momentum is conserved, 
while it remains finite in dimension 3. 
So this is the only explicitely solvable
model that  has a behavior qualitatively consistent with numerical
simulations.

We consider a system of harmonic (linear) coupled oscillators where
the Hamiltonian dynamics are perturbed by a random exchange of momentum
between nearest neighbor atoms. The random exchange of momentum
conserves total momentum and total energy. We construct this
noise with a diffusion on the surface  of constant kinetic energy and
momentum.  Because of the conservation laws, this noise
introduces a certain nonlinearity in the model. 

We compute explicitly the time correlation
function at equilibrium of the energy currents $C_J(t)$ (cf. equation
(\ref{eq:expli})) and we find
that, as $t\sim + \infty$, 
 $C_J(t) \ \mathop{\sim}  t^{-d/2}$ if the system
is unpinned, while  $ C_J(t) \ \mathop{\sim}  t^{-d/2-1}$ 
 if an on-site potential is present. Conductivity, defined by
 Green-Kubo formula, is then finite only in dimension $d\ge 3$ or for
 the pinned system. This indicates a divergence of the conductivity of
 the finite system $\kappa_N$ as $N^{1/2}$ in the unpinned
 one-dimensional case, and as $\log N$ in the unpinned two-dimensional case.

Other explicitly solvable models have been proposed before as
perturbation of the harmonic chain (in \cite{brv, bll} only
the number of particles is conserved, and in \cite{bo} only energy and
the number of particles are conserved). 
In all these models, conductivity is always finite.


In order to compute the conductivity by the Green-Kubo formula,  we
consider the dynamics of the closed system of length $N$ with 
periodic boundary conditions. 
The Hamiltonian is given by
\begin{equation}
  \label{eq:hamilt}
  \mathcal H_N = \frac 12 \sum_{\bx}  \left[ {\bp_\bx^2} +
\bq_\bx \cdot (\nu I - \alpha\Delta) \bq_\bx \right] . \nonumber
\end{equation}
The atoms are labeled by $\bx \in \mathbb T_N^d$, the d-dimensional
discrete torus of length $N$. We denote with $\nabla$ and $\Delta$
respectively the discrete gradient and the discrete Laplacian  on
$\mathbb T_N^d$.
$\{\bq_\bx\}$ are the displacements of the atoms 
from their equilibrium positions.
 The parameter $\alpha > 0$ is the strength of the
interparticles springs, and $\nu \ge 0$ is the strength of the pinning
(on-site potential).

We consider stochastic dynamics where the probability density
distribution on the phase space at time $t$, denoted by  $P(t, \bq,
\bp)$, evolves following the Fokker-Planck equation (cf. \cite{kmh})
\begin{equation*}
 \frac {\partial P}{\partial t} = (- A + \gamma S) P =  L^* P\ .
\end{equation*}
Here $L= A + \gamma S$ is the generator of the process, $L^*$ the
adjoint operator, $A$ is the usual Hamiltonian vector field
 \begin{equation*}
  \label{eq:Agen}
  \begin{split}
    A =& \sum_\bx \left\{ \bp_\bx \cdot \partial_{\bq_\bx} -
     [ (\nu I -\alpha\Delta )
    \bq_\bx] \cdot  \partial_{\bp_\bx} \right\}%
\end{split}
\end{equation*}
while $S$ is the generator of the stochastic perturbation and 
$\gamma > 0$ is a positive parameter that regulates its strength.
The operator $S$ acts only on the momentums $\{\bp_\bx\}$ and
generates a diffusion on the surface of constant kinetic energy and
constant momentum. This is defined as follows. For every nearest
neighbor atoms $\bx$ and $\bz$, consider the ($d-1$)-dimensional surface 
of constant kinetic energy and momentum
\begin{equation*}
  \mathbb S_{e,\bp} = \left\{(\bp_\bx,\bp_\bz)\in \mathbb R^{2d}:
    \bp_\bx^2 + \bp_\bz^2 = 2e ; \; 
    \bp_\bx + \bp_\bz = \bp  \right\} .
\end{equation*}
The following vector fields are tangent to $ \mathbb S_{e,\bp}$
\begin{equation*}
  \label{eq:Xfield}
   X^{i,j}_{\bx, \bz} = (p^j_\bz-p^j_\bx) (\partial_{p^i_\bz}
   - \partial_{p^i_\bx})  -(p^i_\bz-p^i_\bx) (\partial_{p^j_\bz}
   - \partial_{p^j_\bx}) .
\end{equation*}
so $ \sum_{i,j =1}^d (X^{i,j}_{\bx, \bz})^2$ generates a diffusion on
$ \mathbb S_{e,\bp}$. 
In  $d\ge 2$ we define
\begin{equation*}
  \label{eq:Sgen}\begin{array}{ll}
  S & = \displaystyle\frac 1{2(d-1)} \sum_{\bx}\sum_{i,j,k}^d
  \left( X^{i,j}_{\bx, \bx+\be_k}\right)^2
\end{array}
\end{equation*}
where  ${\be}_1,\ldots,{\be}_d$ is the canonical basis of ${\mathbb Z}^d$.
Observe that this noise conserves the total momentum $\sum_\bx
\bp_\bx$ and energy $\mathcal H_N$, i.e.
\begin{equation*}
  S \; \sum_\bx \bp_\bx = 0\ ,\quad S\; \mathcal H_N = 0
\end{equation*}

In dimension 1, in order to conserve total momentum and total kinetic energy, 
we have to consider a random exchange of momentum between three 
consecutive atoms, and we define
$S = \frac 16 \sum_{x\in\mathbb{T}^1_N}(Y_x)^2$, 
where
\begin{equation*}
  \begin{split}
    Y_x\ =\ &(p_x-p_{x+1})\partial_{p_{x-1}}+(p_{x+1}-p_{x-1})\partial_{p_x}\\
    & + (p_{x-1}-p_x)\partial_{p_{x+1}}
  \end{split}
\end{equation*}
which is vector field tangent to the surface of constant energy and
momentum of the three particles involved.

These dynamics can also be written in terms of the solutions of the
stochastic differential equations
\begin{equation}
  \label{eq:sde}
  \begin{split}
    d\bp_\bx = -\left(\nu I - \Delta\right) \bq_\bx \; dt + 2\gamma \Delta
    \bp_\bx \; dt + \sqrt \gamma\; d\bn_\bx(t)
  \end{split}
\end{equation}
where of course $\dot\bq_\bx = \bp_\bx$ and $\bn_\bx(t)$ are defined by
the Ito's stochastic integrals
\begin{equation*}
  \bn_\bx(t) = \frac 1{2\sqrt{d-1}} \sum_{\|\by- \bx\| = 1}
  \sum_{i,j}^d \int_0^t 
  \left(X^{i,j}_{\bx,\by} \bp_\bx\right)(s) \; dw^{i,j}_{\bx,\by}(s)
\end{equation*}
Here $w^{i,j}_{\bx,\by}(t)=w^{i,j}_{\by,\bx}(t) $ are independent
standard Wiener processes. In $d=1$ the expression is similar with  
 the term $2\gamma\Delta \bp_\bx$ 
replaced by $(\gamma/6) \Delta (4p_x + p_{x+1} + p_{x-1})$.

Defining the energy of the atom $\bx$ as
\begin{equation*}
  \label{eq:energyx}
  e_\bx = \frac 12  \bp_\bx^2 +
\cfrac{\alpha}{4}\sum_{\by: |\by -\bx|=1}
(\bq_{\by} - \bq_\bx)^2 + \frac{\nu}2 \bq_\bx^2\ ,
\end{equation*}
the energy conservation law can be read locally as
\begin{equation*}
   e_\bx(t) -   e_\bx(0) =  \sum_{k=1}^d \left(
J_{\bx-\be_k, \bx}(t) - J_{\bx,\bx +\be_k}(t)\right)
 \end{equation*}
where $J_{\bx,\bx +\be_k}(t)$ is the total energy current
between $\bx$ and $\bx +\be_k$ up to
time $t$. This can be written as
\begin{equation}
  \label{eq:tc}
  J_{\bx, \bx +\be_k}(t) = \int_0^t j_{\bx, \bx +\be_k}(s) \; ds +
  M_{\bx, \bx +\be_k}(t) \ .
\end{equation}
In the above $M_{\bx, \bx +\be_k}(t)$ is  the Ito's stochastic integral defined by
\begin{equation*}
   M_{\bx, \bx +\be_k}(t) = \sqrt{\frac{\gamma}{d-1}} \sum_{i,j}^d \int_0^t
  \left(X^{i,j}_{\bx,\bx +\be_k} e_\bx\right)(s) \; dw^{i,j}_{\bx,\by}(s)
\end{equation*}
The instantaneous energy currents $j_{\bx,\bx +\be_k}$ 
satisfy the equation 
\begin{equation*}
  L e_\bx = \sum_{k=1}^d \left(
j_{\bx-\be_k, \bx} - j_{\bx,\bx +\be_k}\right)
\end{equation*}
and can be written as
\begin{equation}
  \label{eq:1}
  j_{\bx, \bx +\be_k} = j^{a}_{\bx, \bx +
\be_k} +\gamma j_{\bx, \bx +\be_k}^s \quad .
\end{equation}
The first term in (\ref{eq:1}) is the Hamiltonian contribution to the
energy current
\begin{equation}
  \label{eq:2}
    j^a_{\bx,\bx +\be_k} = -\cfrac{\alpha}{2} (\bq_{\bx+\be_k} -
\bq_\bx)\cdot (\bp_{\bx+\be_k} + \bp_\bx)
\end{equation}
while the noise contribution in $d\ge 2$ is
\begin{equation}
  \label{eq:3}
     \gamma j^s_{\bx,\bx +\be_k} =- \gamma\nabla_{\be_k} \bp_\bx^2
 \end{equation}
and in $d=1$ is
\begin{equation*}
\begin{split}
  \gamma j^s_{x, x + 1} &= -\gamma\nabla \varphi (p_{x-1},p_x,p_{x+1})
  \\
 \varphi (p_{x-1},p_x,p_{x+1})=& 
\frac 16 [p_{x+1}^2 + 4 p_{x}^2 +
    p_{x-1}^2 +  p_{x+1} p_{x-1}\\
    & -2 p_{x+1} p_{x}-2 p_{x} p_{x-1}]
\end{split}
\end{equation*}

Consider the dynamics of the closed system on $\mathbb T^d_N$ in
microcanonical equilibrium. 
The microcanonical distribution is usually defined as the uniform measure 
on the energy surface $\mathcal H = N^d e$, for a given
$e>0$. 
Our dynamics conserve also $(\sum \bp_\bx)^2 + \nu (\sum
\bq_\bx)^2$.
Notice that the dynamics is invariant under the change of coordinates
  $\bp_\bx'= \bp_\bx- \sum_\by \bp_\by$ and  $\bq_\bx'= \bq_\bx-
  \sum_\by \bq_\by$.  Consequently,
without any loss of generality, we can fix
 $\sum \bp_\bx = 0$ and $\sum \bq_\bx= 0$ in the microcanonical
 measure. 

Let us define $\frak J_{\be_1} = \sum_\bx j_{\bx, \bx+\be_1} =
\sum_\bx j^a_{\bx, \bx+\be_1}$. We are interested in computing the
correlation function:
\begin{equation}
  \label{eq:8}
  C_{1,1} (t) = \lim_{N\to \infty} \frac 1{N^d} \mathbb E(\frak J_{\be_1}(t)\frak
  J_{\be_1}(0)) 
\end{equation}
where $\mathbb E$ is the expectation starting with the
microcanonical distribution defined above.
By explicit calculation we can solve the equation
\begin{equation}
  \label{eq:u}
   \left(\lambda - L\right)^{-1}\frak
      J_{\be_1} = - \frac \alpha\gamma \sum_{\bx,\by}
 g_{\lambda,N}(\bx - \by) \bp_\bx\cdot \bq_{\by} 
\end{equation}
where $g_{\lambda,N}(\bx)$ is 
the solution of the equation 
\begin{equation}
  \label{eq:g}
  \frac{2\lambda}{\gamma} g_{\lambda,N}(\bx) - 4\Delta g_{\lambda,N}
  (\bx) =  (\delta(\bx +\be_1) - \delta(\bx - \be_1) )
\end{equation}
on $\mathbb T_d^N$ for $d\ge 2$, while in one dimension it solves
\begin{widetext}
  \begin{equation}
    \label{eq:g1d}
        \frac{2\lambda}{\gamma} g_{\lambda,N}(x) - \frac 13\Delta\left[
        4 g_{\lambda,N} (x) + g_{\lambda,N}(x+1) + g_{\lambda,N} (x-1)\right] 
      = (\delta(x + 1) - \delta(x - 1) ) .
    \end{equation}
\end{widetext}
Consequently, for $\lambda>0$, 
 we can write the Laplace transform of $C_{1,1}(t)$ as
\begin{equation}
  \label{eq:11}
  \int_0^\infty e^{-\lambda t} C_{1,1}(t) \; dt = \lim_{N\to\infty}
  \left<j^a_{0,\be_1} \left(\lambda -  L\right)^{-1} \frak
      J_{\be_1}  \right>_N 
\end{equation}
where $<\cdots>_N$ denotes the expectation with respect to the
microcanonical measure.

Substituting (\ref{eq:u}) in (\ref{eq:11}) and using equivalence of
ensembles estimates   (see \cite{bbo} for the details) we have
\begin{equation}
  \label{eq:88}
  \begin{split}
    \int_0^\infty & e^{-\lambda t} C_{1,1}(t) \; dt \\
     =&\  \frac {\alpha^2
      e^2}{2 d \gamma} \sum_\bz g_\lambda(\bz) \left[\Gamma(0,\bz+
      \be_1) - \Gamma(0,\bz - \be_1)\right]
  \end{split}
\end{equation}
where $\Gamma$ is the kernel of the operator $(\nu I -
\alpha \Delta)^{-1}$ on $\mathbb Z^d$, while $g_\lambda$ is the solution of
equations (\ref{eq:g}) in $\mathbb Z^d$ or (\ref{eq:g1d}) in $\mathbb Z$. 
We compute explicitly (\ref{eq:88}) and invert the Laplace trasform
obtaining
\begin{equation}
  \label{eq:expli}
  C_{1,1}(t) = \frac{ e^2}{4\pi^2 d} \int_{[0,1]^d} (\partial _{k^1}\omega(\bk))^2
  \; e^{-t\gamma \psi(\bk)} \;  d\bk
\end{equation}
where $\omega(\bk) = ({\nu + 4\alpha  \sum_{j=1}^d \sin^2(\pi
  k^j)})^{1/2}$ is the dispersion relation of the system, and   
\begin{equation}
  \label{eq:psi}
  \psi(\bk) =  \ \left\{
\begin{array}{ll}
8 \sum_{j=1}^d \sin^2(\pi k^j)  & \hbox{if $d\ge 2$}\; , \\
4/3 \sin^2(\pi k)(1+ 2 \cos^2(\pi k)) & \hbox{if $d = 1$}\; 
\end{array}
\right.
\end{equation}
Since around $\bk = 0$ we have $\psi(\bk) \sim \bk^2$ and \\
$(\partial _{k^1}\omega(\bk))^2 \sim (\alpha k^1)^2 (\nu + \alpha^2 \bk^2)^{-1}$,
 we have the following asymptotic behavior
\begin{equation}
  \label{eq:4}
  C_{1,1}(t) \ \mathop{\sim}_{t\to\infty} \ \left\{
\begin{array}{ll}
t^{-d/2}  & \hbox{if $\nu = 0$}\; , \\
 t^{-d/2-1} & \hbox{if $\nu > 0$}\; 
\end{array}
\right.
\end{equation}

 By Green-Kubo formula \cite{kmh}, the
conductivity in the direction $\be_1$ is given by
\begin{equation}
  \label{eq:9}
 \begin{split}
  \kappa^{1,1} = \lim_{t\to\infty} \frac 1{2 e^2 t} \lim_{N\to \infty} \sum_\bx
    \mathbb E\left(J_{\bx,\bx+\be_1}(t) J_{0,\be_1}(t) \right)
 \end{split}
\end{equation}

By explicit calculation one can show (see \cite{bbo} for the details)
\begin{equation}
  \label{eq:10}
  \begin{split}
    \kappa^{1,1} &= \frac{\gamma}d + \frac 1{2 e^2} \int_0^\infty
    C_{1,1}(t) \; dt \\
    &= \frac{\gamma}d +  \frac 1{8\pi^2 d \gamma} 
    \int_{[0,1]^d} \frac{(\partial _{k^1}\omega(\bk))^2}{\psi(\bk)} \;  d\bk.
  \end{split}
\end{equation}
By (\ref{eq:4}), if $d\ge3$ or if $\nu >0$, the integral on the right
hand side of (\ref{eq:10}) is finite.

If $\nu = 0$ and $d\le 2$, by  (\ref{eq:4}) the time integral in
(\ref{eq:10}) diverge and conductivity is infinite. Following
Ref. \cite{sll} (page 46), one can estimate the dependence of
the conductivity of the finite open system of length $N$ with thermic
baths at the boundary imposing a temperature gradient, 
by restricting the time integral in (\ref{eq:10}) to times
smaller than the ''transit time`` $N/v_s$, where $v_s$ is the sound
velocity of the lattice defined as $v_s = \lim_{\bk\to
  0} \partial_{k^1}\omega(\bk) = 1$. 
This gives a finite size
conductivity $\kappa_N$ diverging like $N^{1/2}$ in dimension 1, and
like $\log N$ in dimension 2.

\medskip

\emph{Discussion}. 
The exact results presented in this letter concerning the stochastic
model we introduced give some indications about the role of
conservation of momentum and of confinement (pinning) in heat
conduction for the nonlinear deterministic Hamiltonian case. In fact
the decay of the energy current correlations and consequently the 
  behavior of the conductivity that we proved in our stochastic model are
(qualitatively)  the same as those indicated by numerical simulation
for the deterministic non-linear FPU models.
Furthermore a recent paper   \cite{lng} on the one dimensional unpinned purely
quartic FPU model suggest 
the same decay of the time-correlations of energy current 
as in our stochastic model. 

In the one-dimensional case we can give the following euristic
explanation of the effect of the noise in these harmonic systems. 
In deterministic harmonic systems the energy of each mode is conserved, 
in both pinned or unpinned chain; so if modes are created by initial
or boundary conditions, they cannot interact and they move
ballistically.
This causes ballistic transport and diverging conductivity, in both
cases (cf. \cite{rll, pc}). 
The effect of the energy-momentum conservative noise we have
introduced is to scatter  modes randomly with rate proportional to
$k^2$, for small wave number $k$.
The  \emph{velocity}  of the $k$ mode
 is given by the gradient of the dispersion
function $\nabla\omega(k)$.
 In the unpinned chain $\nabla\omega(k) \sim 1$ for small $k$, 
 so  small wave number modes have little probability to be
scattered, and their movement results in  a ballistic contribution to energy
transport, while modes with large $k$ scatter fast and consequently they
diffuse. Properly averaging over all modes one obtains a current proportional
to $N^{-1/2}$, i.e. a conductivity diverging like $N^{1/2}$.
  In the pinned chain small
wavenumber modes move very slowly [$\nabla\omega(k) \sim k$], so
there is a  high probability they will be  scattered and then diffuse while 
they cross the system. Consequently in this case conductivity is
finite. 

In \cite{bo} we considered
 the unpinned 1-dimensional harmonic chain with noise that conserves
 only  energy, and prove that conductivity is finite in any
 dimension. In this last case all modes are scattered with constant
 rate (independent of $k$). 

In non-linear FPU type of interaction, a
behavior  $\kappa_N \sim N^{\alpha}$, with $0< \alpha <1$ is 
observed numerically. 
 But numerical simulations are not conclusive about
the value of  
$\alpha$ and there is an intense debate in the literature on
this value (cf. \cite{sll}). As suggested recently by
 Livi \cite{livi},
in one-dimensional
systems 
the value of $\alpha$ may depend on the specific non-linearity
of the interaction (unlike the logarithmic behavior of the two-dimensional
        systems).
The non-linearity creates some scattering of the longwave
modes, which results in a breaking of the ballistic transport, and in
a superdiffusive behavior of these modes. An extreme case is given by
the 1-dimensional coupled-rotors model, which is an example of a
non-linear chain that conserves momentum 
and has finite conductivity \cite{glpv}.  In this example, 
 rotobreathers (isolated rotors with high kinetic energy
that turn very fast) scatter waves that try to pass through them 
(cf. \cite{fmf}).

We thank Joel Lebowitz, Roberto Livi and  Herbert Spohn for their
interest in this work and very useful discussions.
We acknowledge the support of the ACI-NIM 168
\emph{Transport Hors Equilibre} of the Minist\`ere de l'Education
Nationale, France.

\end{document}